\documentclass[11pt,a4paper]{article}

\usepackage{jheppub}
\usepackage{epsfig}
\usepackage{graphicx}
\usepackage{color}
\usepackage{latexsym}
\usepackage{amssymb}
\usepackage{slashed}
\usepackage{dcolumn}

\def\nn{\nonumber}

\def\MeV{\textrm{MeV}}
\def\GeV{\textrm{GeV}}

\def\pp{p^{\prime}}
\def\qp{q^{\prime}}
\def\ub{\bar{u}}
\def\qb{\bar{q}}
\def\lt{l_{\bot}}

\def\Q8{Q_{8g}}

\def\Slal{\slashed{l}}
\def\Slap{\slashed{p}}
\def\Slaq{\slashed{q}}

\def\gammu{\gamma_{\mu}}
\def\gamnu{\gamma_{\nu}}
\def\gam5{\gamma_5}

\title{ \bf
Order $\alpha_s^2$ magnetic penguin correction \\
for $B$ decay to light mesons
}

\author[a]{C.~S.~Kim}
\author[b]{and Yeo Woong Yoon}

\affiliation[a]{Department
  of Physics and IPAP, Yonsei University, Seoul 120-749, Korea}
\affiliation[b]{School of Physics, KIAS, Seoul 130-722, Korea}

\emailAdd{cskim@yonsei.ac.kr}
\emailAdd{ywyoon@kias.re.kr}

 \abstract{
 We  compute the order $\alpha_s^2$ correction to the matrix element of magnetic penguin
 operator  for $B$ meson decaying to light mesons within the
 QCD factorization framework.  We explicitly show that the soft and collinear divergences are
 canceled out,  so that the validity of QCD factorization is confirmed.
 We present the result of the calculation in complete analytic forms.
 The result is also applied to $B \rightarrow K \pi$ decays, and we find that the order $\alpha_s^2$
 correction of magnetic penguin operator can considerably reduce the
 coefficient of penguin amplitude $a^c_{4,I}$. The reduction is stronger for the imaginary part.
 }


 \arxivnumber{1107.1601}
 \begin{document}
 \maketitle
 \flushbottom

 \section{Introduction}

  Non-leptonic $B$ decays have great affluence of phenomenological applications. Especially,
  they provide a plenty of fascinating CP asymmetries in the standard model (SM), where we have a number
  of chances   to test the SM by searching discrepancies between the theoretical predictions and
  experimental measurements. The measurements of CP asymmetries have been performed well in various
  $B$ factories such as Belle and BaBar and also in Tevatron. Furthermore, upcoming $B$ factories
  such as LHC-b, Super-B and Belle-2 will shed more light on CP asymmetries by accumulating much
  larger amount of data.

  For a long time, $B \to K \pi$ decays have been a continuously engrossing issue due to that some
  puzzling behavior occurs  in their CP asymmetry measurements: The difference between
  $A_{CP}(\pi^\mp K^\pm)$ and $A_{CP}(\pi^0 K^\pm)$ measured by Belle, BaBar and CDF collaborations
  is unexpectedly larger than the theoretical estimates
  \cite{:2008zza,Aubert:2007hh,Aubert:2007mj,Aaltonen:2011qt} in the SM. Recent brand-new
  result from the LHC-b experiment also supports the behavior \cite{Perazzini:2011kd}. There have
  been numerous new physics analyses on the issue model-independently or model-dependently.
  However, before we explore new physics effects beyond the SM to resolve the puzzle, it is strongly
  required that one should perform more precise higher order corrections to the observables within
  the SM.

  The QCD factorization (QCDF) is a theoretical framework for systematically calculating hadronic
  matrix element of weak decays of $B$ meson \cite{Beneke:1999br}.  It has been extensively applied to
  non-leptonic $B$ decays \cite{Beneke:2000ry,Beneke:2001ev,Beneke:2003zv}.
  The QCDF formalism for hadronic matrix element of non-leptonic $B$ decays is described by
  \begin{eqnarray}
  \label{QCDF-framework}
&&  \langle M_1 M_2 | Q_i | B \rangle =
  \sum_j F^{B\to M_1}_j(m_2^2) f_{M_2} \int^1_0 du T^I_{ij}(u) \Phi_{M_2}(u)
  +(M_1 \leftrightarrow M_2) \nn \\
&&~~~~~  +f_B f_{M_1} f_{M_2}
 \int^1_0 d\xi du dv  T^{II}_{i}(\xi,u,v) \Phi_B(\xi) \Phi_{M_1}(v) \Phi_{M_2}(u)\, ,~~~~~~~~~~
  \end{eqnarray}
  where $F^{B\to M_i}$ is $B \to M_i$ form-factor, $f_{M_i}$ is decay constant of meson $M_i$,
  and $\Phi_{M_i}(u)$ is light-cone distribution amplitude of meson $M_i$ with parton
  momentum fraction $u$. The formalism is expressed by convolution of hard-scattering kernel
   $T^{I,II}$ with meson distribution amplitude at the leading power of $\Lambda_{QCD}/m_b$.
  The hard-scattering kernel is separated into the hard-scattering  form-factor term $T^I$
  and the hard-scattering spectator term $T^{II}$.
  The very first paper of QCDF calculated next-to-leading order (NLO) correction to
  $B \to \pi\pi$ decays \cite{Beneke:1999br}. The imaginary part of the decay amplitude,
  which causes strong CP phase, arises first at the order $\alpha_s$ from hard-scattering
  contribution between decaying quarks of different mesons. Therefore, the next-to-next-to
  leading order (NNLO) correction is of great importance in perturbative expansion in order to
  provide reliable prediction for CP asymmetries.

   The NNLO correction to tree amplitude for hard-scattering form-factor term has
   been calculated first for imaginary part \cite{Bell:2007tv}  and later  for real part
   \cite{Bell:2009nk}, where the order $\alpha_s^2$ contribution is quite significant
   compared to the order $\alpha_s$ contribution. Especially, the color-suppressed amplitude is very
   sensitive to the order $\alpha_s^2$ contribution. The result is also confirmed by ref.~\cite{Beneke:2009ek}.
   As for the hard-scattering spectator term, the one-loop order $\alpha_s^2$ correction has been
   calculated for tree amplitude \cite{Beneke:2005vv,Pilipp:2007mg,Kivel:2006xc} and for
   penguin amplitude \cite{Beneke:2006mk,Jain:2007dy}.
   However, the order $\alpha_s^2$ penguin correction for the hard-scattering form-factor term has
   not yet been calculated.
   This correction is more important for $B \to K \pi$ decays because
   the tree amplitudes are CKM-suppressed and the decays are penguin-dominant.
   Especially, the order $\alpha_s^2$ penguin correction is highly required for theoretical
   estimate of CP asymmetries since it is the first correction in perturbative expansion.

    Motivated by current status of higher order correction for  $B$ to light
   meson decays,
   here we compute the order $\alpha_s^2$ one-loop contribution (NLO) for the
    magnetic penguin operator as the first step
    toward the complete order $\alpha_s^2$ correction of penguin amplitude.
   In section 2, we describe the formalism for the calculation.
   In order to regularize ultra-violet (UV) divergences and infra-red (IR) divergences in the
   NLO calculation, we use dimensional regularization where the space-time dimension $d$ is
   analytically continued to $d = 4-2 \,\varepsilon$. Especially,
   we choose naive dimensional regularization (NDR) scheme for the prescription of $\gamma_5$.
   We use 't Hooft-Feynman gauge ($\xi=1$) throughout the calculation.
    In section 3, we explicitly show that the soft and collinear divergences
   are canceled out in the calculation.
   The result of calculation is expressed in section 4 in complete analytic forms.
   We relegate our analytic results of the master integrals to appendix.
   We apply our result to $B \to K\pi$ decays and provide numerical values of  the NLO correction for
   magnetic penguin operator
   compared with other contributions. In section 5, we summarize and conclude.

\section{Formalism}

 The effective amplitude of non-leptonic weak decays of $B$ meson, such as
 $B \to M_1 M_2$, is expressed  \cite{Buchalla:1995vs} by
  \begin{equation}
 A_{eff} = \frac{G_F}{\sqrt{2}} \sum_i \lambda^i \,
  C_i(\mu) \langle M_1 M_2 | Q_i | B \rangle (\mu) \,
 \end{equation}
 with effective operators $Q_i$, $i =1,2,...,10, 7\gamma, 8g$. $G_F$ is Fermi constant and $\lambda^i$ is
 a factor for CKM matrix elements corresponding to operator $Q_i$.
 The magnetic penguin operator $\Q8$, on which we
 are focusing in this work, is defined by
 \begin{equation}
 \Q8 = -\frac{g_s}{8\pi^2} m_b \bar{s} \sigma_{\mu\nu} ( 1+\gam5) G^{\mu\nu} b \,,
 \end{equation}
 where $G^{\mu\nu}$ is gluonic field strength tensor contracted by SU(3) generators.
 We do not consider the magnetic $\gamma$-penguin operator $Q_{7\gamma}$ because the contribution is
 much suppressed by the fine-structure constant.
 The Wilson coefficients $C_i(\mu)$ are to be calculated near weak boson mass scale $M_W$ where we
 can safely avoid large logarithms in perturbative expansion. Then, $C_i(\mu)$  are evolved into
  $m_b$ scale through renormalization group equation associated with anomalous dimension matrix.
 In principle, the  scale dependence of $C_i(\mu)$ should be canceled by the scale dependence
  of the full calculation of matrix element $\langle M_1 M_2 | Q_i | B \rangle (\mu)$
  up to the order of truncation of perturbative expansion.

\begin{figure}[t]
\centerline{\epsfig{figure=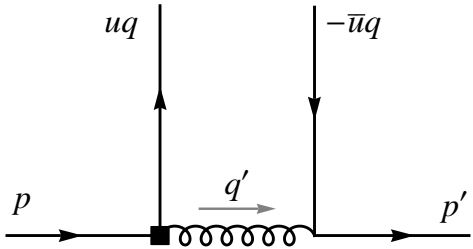, scale=0.5}}
\caption{Order $\alpha_s$ (LO) diagram for magnetic penguin operator with momentum configuration.
$\ub = 1-u$ is understood. The square box denotes $\Q8$ insertion.}
\label{fig-LO}
\end{figure}

\begin{figure}[t]
\centerline{\epsfig{figure=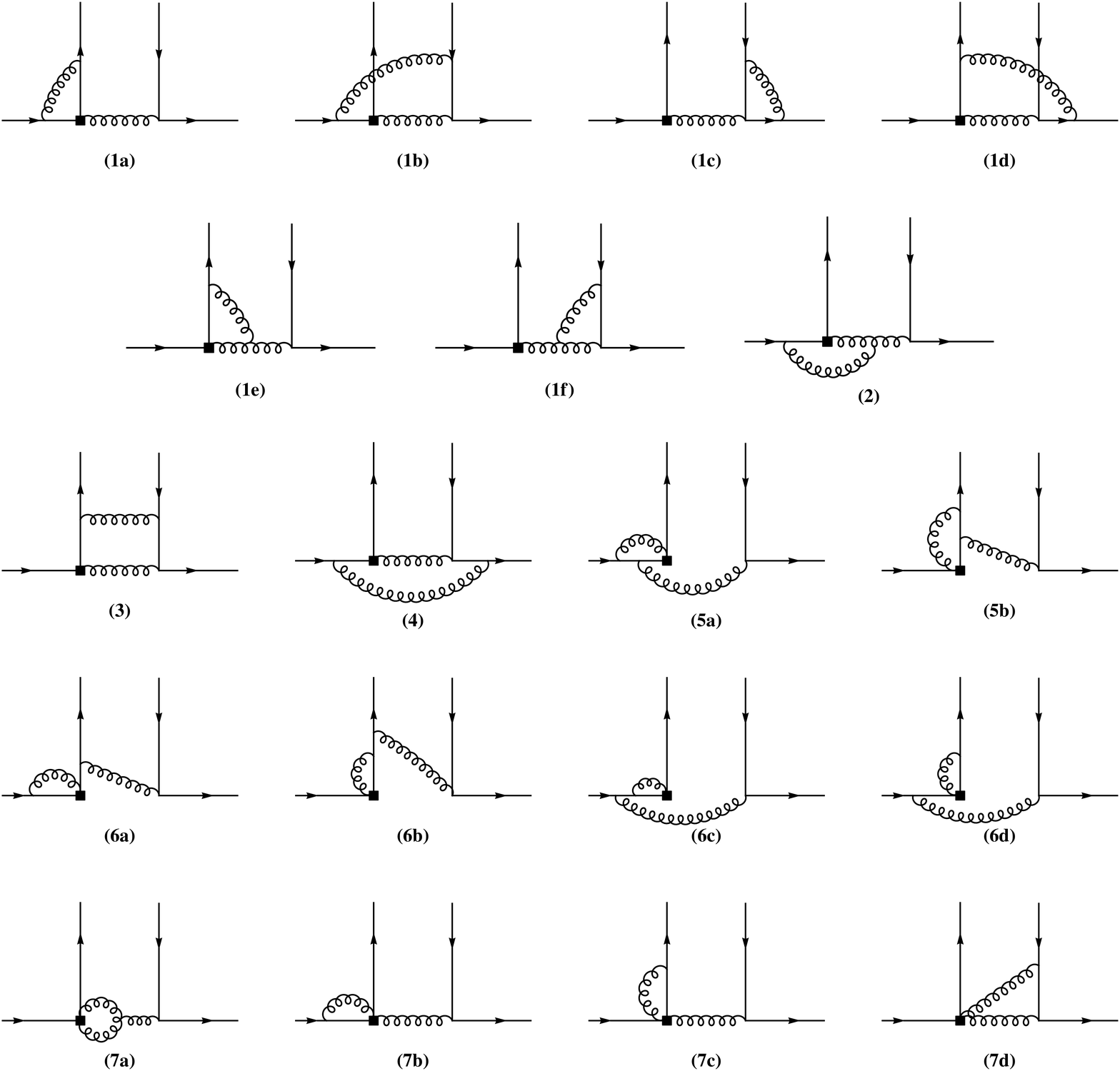, scale=0.28}}
\caption{NLO diagrams for magnetic penguin operator. The momentum configurations for all the diagrams
 are same as in figure~\ref{fig-LO}. The square box denotes $\Q8$ insertion.}
\label{fig-NLO}
\end{figure}

  The calculation of hadronic matrix element can be handled within the
  QCDF framework. The formalism is expressed in eq.~(\ref{QCDF-framework}).
   In this framework soft gluon exchange between decaying quarks is power suppressed in heavy quark limit.
   The dominant hard gluon scattering contribution is absorbed into hard-scattering kernels
   $ T^{I, II}$.
   The matrix element is described by convolution of hard-scattering kernel with meson
   distribution amplitude.
   Basically, $T^{I}$ starts at order unity in perturbative
  expansion, while $ T^{II}$ starts
  at order $\alpha_s$. It should be noted that the contribution of magnetic penguin operator
  in $T^{I}$ starts at order $\alpha_s$. In order word, leading order (LO) diagram for
  magnetic penguin operator is order $\alpha_s$ as shown in figure~\ref{fig-LO}.
  This property  is same with all other penguin amplitudes. In the figure~\ref{fig-LO},
  spectator quark line is suppressed.
  Left horizontal line denotes incoming $b$ quark with momentum $p$ and right horizontal line is
  outgoing quark with momentum $p'$ which forms a bound state of a meson with spectator anti-quark.
  The upper side lines represent the quark and anti-quark that make a bound state of another meson
  with momentum $q$.
  $u$ and $\ub$ (equal to $1-u$) is parton momentum fraction of each quark in the upper side meson.
  The kinematic relations among the momentum four vectors are as follow:
 \begin{eqnarray}
 p^2 = m_b^2,~~q^2 = 0,~~p'^2=0,~~ q'^2 = {\bar u} m_b^2,
 ~~~~~~~~~~~~~~~~~~~\nn\\
 p\cdot q = p\cdot p'= p'\cdot q= q\cdot q' = \frac{m_b^2}{2},
  ~~ p'\cdot q' = \frac{{\bar u} m_b^2}{2},
  ~~ p \cdot q' = \frac{(1+\bar u) m_b^2}{2}\,.
 \end{eqnarray}
  In order to obtain hard-scattering kernel $T$ (superscript $I$ is suppressed) for
  magnetic penguin operator, we use the QCDF formula which is schematically described by
 \begin{equation}
 \label{eq:match0}
 \langle \Q8 \rangle_{\rm{ren}} = F \cdot T \otimes \Phi_M\,,
 \end{equation}
where the symbol $\otimes$ denotes convolution with meson distribution amplitude.
 $\langle \Q8 \rangle_{\rm{ren}}$ is renormalized matrix element which is related with
 its bare quantity by
 \begin{equation}
 \label{eq:OPRen}
\langle \Q8 \rangle_{\rm{ren}} = Z_m Z_g Z_q Z^{1/2}_G { Z}_{88} \langle \Q8 \rangle_0\,,
\end{equation}
where ${Z}_{88}$ is operator renormalization constant for $\Q8$, and $Z_m,~Z_g,~ Z_q,~ Z_G$ are
 the mass, coupling, quark field and gluon field renormalization constant, respectively.
 The one-loop calculation of ${Z}_{88}$ is shown in ref.~\cite{Grinstein:1990tj}.
 We summarize each value of one-loop renormalization constant:
 \begin{eqnarray}
 Z_{m}^{(1,1)} &=& -3C_F \,,\nn \\
 Z_g^{(1,1)} &=& -\frac{11}{6} N_c + \frac{1}{3}n_f \,,\nn \\
 Z_q^{(1,1)} &=& -C_F \,,\nn \\
  { Z}_{88}^{(1,1)} &=& 8C_F - 2N_c \,,
 \end{eqnarray}
 where the following power expansion is considered
 \begin{equation}
 Z_i = 1+ \sum^\infty_{j=1} \sum^j_{k=1} \left( \frac{\alpha_s}{4\pi} \right)^j
\frac{1}{\varepsilon^k}
 Z_{i}^{(j,k)} \,.
 \end{equation}
 Each quantity in the factorization formula eq.~(\ref{eq:match0}) can be power expanded.
 Since $\langle \Q8 \rangle_{\rm{ren}}$ as well as $T$ start at order $\alpha_s$, we describe
 \begin{eqnarray}
 \langle \Q8 \rangle_{\rm{ren}} &=& \left( \frac{\alpha_s}{4\pi} \right)
 \left( \langle \Q8 \rangle_{\rm{ren}}^{(0)} +
   \left( \frac{\alpha_s}{4\pi} \right) \langle \Q8 \rangle_{\rm{ren}}^{(1)} +
   {\cal O}(\alpha_s^2) \right) \,,\nn \\
 F &=& F^{(0)} +  \left( \frac{\alpha_s}{4\pi} \right) F^{(1)} + {\cal O}(\alpha_s^2) \,,\nn \\
 \Phi_M &=& \Phi^{(0)}_M +  \left( \frac{\alpha_s}{4\pi} \right) \Phi^{(1)}_M + {\cal O}(\alpha_s^2) \,,\nn \\
  T &=& \left( \frac{\alpha_s}{4\pi} \right) \left(
T^{(0)} + \left( \frac{\alpha_s}{4\pi} \right) T^{(1)} + {\cal O}(\alpha_s^2) \right) \,.
 \end{eqnarray}
 From the power expansion of factorization formula eq.~(\ref{eq:match0}), we obtain
 \begin{eqnarray}
 \label{eq:master}
 \langle \Q8 \rangle_{\rm{ren}}^{(0)} &=&  T^{(0)}\,, \\
 \label{eq:match}
 \langle \Q8 \rangle_{\rm{ren}}^{(1)} &=& F^{(0)}T^{(1)} \otimes \Phi^{(0)}_M
 + F^{(1)}T^{(0)} \otimes \Phi^{(0)}_M + F^{(0)}T^{(0)} \otimes \Phi^{(1)}_M \,.
 \end{eqnarray}
 Those eqs.~(\ref{eq:master},\ref{eq:match}) are the master formula for the calculating hard-scattering kernels.
 In those formula, all the quantities are renormalized  while they might have
 IR divergences: All the IR divergences should be canceled in those equations.
 In the next section it will be explicitly shown that the soft and collinear divergences
 are canceled out in those equations.
 The NLO diagrams for magnetic penguin operator are shown in figure~\ref{fig-NLO}.

   In order to compensate scale dependence of NLO hadronic matrix element, we have to employ
   next-to-leading logarithmic (NLL) Wilson coefficients for the $\Q8$.
   We refer to refs.~\cite{Chetyrkin:1996vx,Greub:2000sy}
  for NLL effective Wilson coefficient $C_{8g}^{\rm{eff}}$ at low energy scale.
  The light-cone projection operator of a light pseudoscalar meson in momentum space is applied to
  upper side quark and anti-quark spinors. The projection operator
 in leading twist is given by
\begin{equation}
\label{eq:proj}
M^P_{\alpha\beta} = \frac{i f_P}{4 N_c} \left[ \Slaq \gamma_5 \right]_{\alpha\beta} \Phi_M(u)\,,
\end{equation}
where $\alpha,\beta$ are spinor indices of upper side quark and anti-quark spinors.
 The distribution amplitude of meson $M$ denoted by $\Phi_M(u)$ is conventionally expanded in
 Gegenbauer polynomials:
\begin{equation}
\Phi_M(u,\mu) = 6u(1-u) \left[ 1 + \sum^\infty_{n=1} \alpha_n^M(\mu) C^{(3/2)}_n (2u-1)\right]\,.
\end{equation}
$\alpha_n^M(\mu)$ are Gegenbauer moments and $C^{(3/2)}_n$ are Gegenbauer polynomials of order $\frac{3}{2}$.
 We truncate the expansion by $n=2$. In order to consistently compensate NLO hadronic matrix element,
 we also use NLL evolution equation for Gegenbauer moments
 which are obtained in \cite{Floratos:1977au,GonzalezArroyo:1979df,Mueller:1993hg}:
\begin{eqnarray}
\alpha_n^M(\mu) &=& U_{nn} \alpha_n^M(\mu_0) + \delta_{n,2} \,U_{20}\,, \nn \\
U_{nn} &=& \left( \frac{\alpha_s(\mu_0)}{\alpha_s(\mu)} \right)^{\frac{\gamma_n^{(0)}}{2\beta_0}}
\left[ 1+ \left( \frac{\gamma_n^{(1)}}{2\beta_0} - \frac{\gamma_n^{(0)}\beta_1}{2\beta_0^2}\right)
\frac{\alpha_s(\mu_0)-\alpha_s(\mu)}{4\pi} \right] \,, \nn
\end{eqnarray}
\begin{eqnarray}
U_{20} &=&  \left(\frac{\alpha_s(\mu)}{4\pi}\right)
\Bigg[1-\left( \frac{\alpha_s(\mu_0)}{\alpha_s(\mu)} \right)^{1+\frac{\gamma_2^{(0)}}{2\beta_0}} \Bigg]
\frac{\gamma_2^{(0)}}{\gamma_2^{(0)}+2\beta_0}
~\frac{7}{6}\left(1 - \frac{1}{5}\beta_0\right)\,,
\end{eqnarray}
where the anomalous dimension matrix elements are given by $\gamma_1^{(0)}=-\frac{64}{9}$,
$\gamma_2^{(0)} = -\frac{100}{9}$ for one-loop order, and $\gamma_1^{(1)}=-\frac{15808}{243}$,
$\gamma_2^{(1)} = -\frac{22000}{243}$ for two-loop order. We set $n_f = 5 $ for each number.

\section{Soft, Collinear and UV divergence}

\subsection{ Soft and collinear divergence cancelation }

 The soft divergences can occur when the loop momentum $\l$ goes to 0. The collinear
 divergences can arise when $\l$ becomes collinear to the momentum of the massless quark $q$ or $\pp$.
 All these soft and collinear divergences should be canceled out in order to verify that the
 QCDF framework is valid. In this subsection we explicitly show that the soft and collinear
 divergences of  the order $\alpha_s^2$ correction for magnetic penguin operator are canceled out in
 the QCDF framework.

 First, we examine the soft divergence cancelation. We set the loop momentum $l\sim\lambda$ where
 $\lambda$ is approximately 0 which can be used for investigating the degree of soft divergence.
 In this way we can easily check the soft divergence by counting the power of $\lambda$ in the integral. We note that
  $d^4 l \sim \lambda^4$. If the power of $\lambda$ is smaller than or equal to 0, the integration has soft divergence.

  It turns out that the diagrams (1e), (1f), (2) and (5a), (5b), $\cdots$, (7d) have no soft divergences.
 The soft divergence of the diagram (3) vanishes due to the on-shell condition. The discussion
  about the diagram (4) will be done in the later part of this section. We now consider diagrams
  (1a) and (1b). Except overall factor including color factor (if not mentioned, we imply this
  suppression below), their integrand is
 \begin{eqnarray}
 & & (1a)+(1b) = \frac{1}{l^2 ((l+p)^2-m_b^2)} \bigg( \frac{[\gamma^\nu(\Slal+u \Slaq) \Gamma^\mu_{\qp}
(\Slal+\Slap+m_b)\gamnu][\gammu]}{{\qp}^2 (l+u q)^2}
~~~~~~~~~~~~~~~~~~~~~~~~~~~~~~  \nn \\
 & & ~~~~~~~~~~~~~~~~~~~~~~~~~~~~~~~~~~~~~~~~~~~~~~~~~
 -\frac{[\Gamma^\mu_{l+\qp}(\Slal+\Slap+m_b)\gamma^\nu][\gammu(\Slal+\ub \Slaq)\gamnu]}{(l+\qp)^2 (l+\ub q)^2}  \bigg)\,,
\end{eqnarray}
where we suppress the spinor indices. We use square brackets implying separate spinor indices.
$\Gamma^\mu_{\qp}$ is a Dirac structure coming from $\Q8$ insertion which is defined by
\begin{equation}
\Gamma^\mu_{\qp} = \sigma^{\mu\nu} \qp_\nu(1+\gam5)\,.
\end{equation}
 After $l$ goes to 0 and doing some Dirac algebra, we easily find the terms in the round brackets
 cancel each other. Similarly, one can find that the soft divergences from diagrams (1c) and (1d)
 are also canceled each other. Then, we reach that all the soft divergences are safely canceled.

 For the collinear divergence, we first consider the case where loop momentum $l$ becomes collinear
  to the momentum $q$  of upper side quark lines.  We typically decompose $l$ as follows,
 \begin{equation}
 l=\alpha q + \beta \qb + \lt^2\,,
 \end{equation}
 where $q = (E,0,0,E)$, $\qb = (E,0,0,-E)$ and $\lt=(0,l_1,l_2,0)$. $E$ is energy of upper side decaying meson
 which is approximately $E \sim m_b/2$. Because $l$ is collinear to $q$ we set the order of each parameter
 such that $\alpha \sim 1$, $\beta \sim \lambda^2$ and $\lt^2 \sim \lambda^2 m_b^2$,
 where $\lambda \sim \Lambda_{\rm{QCD}}/m_b$.
 We note that $ d^4l =1/2dl_q dl_{\qb}d\lt^2 \sim \lambda^4 m_b^4$  and $l^2 \sim \lambda^2 m_b^2$.
 Then we count the power of $\lambda$ in the integral which implies the degree of collinear divergence.
 Similar to the soft divergence, if the power of $\lambda$ is less then or equal to 0, the integral has collinear divergence.
 One may simply check the collinear divergences from the fact that the gluon, which is connected any one
 of the external massless quark line, potentially has collinear divergence.

  It is simple to find that the diagrams (2), (5a), (5b), $\cdots$, (7c) have no collinear
   divergence. It is worth showing the integrand of diagram (7d) for studying the collinear
   divergence:
 \begin{equation}
 (7d) = \frac{[\tilde\Gamma^{\mu\nu}][\gamnu\Slal\gammu]}{l^2 (l+\ub q)^2 (l-\pp)^2}\,,
 \end{equation}
 where $\tilde\Gamma^{\mu\nu}$ is the Dirac structure from the $qqgg$ vertex of $\Q8$ which is defined by
 \begin{equation}
 \tilde\Gamma^{\mu\nu} = \sigma^{\mu\nu} (1+\gam5)\,.
 \end{equation}
 After substituting $l$ for $\alpha q$ in the numerator and using some Dirac algebra and on-shell condition,
 we find that the numerator vanishes. Now we consider again diagram (1a) including color factor:
 \begin{equation}
 (1a) =\Big(-\frac{C_F}{2N_c}\Big) \frac{1}{{\qp}^2 l^2 (l+u q)^2} \bigg( [\gamma^\nu(\Slal+u \Slaq) \Gamma^\mu_{\qp}
 \frac{1}{\Slal + \Slap - m_b}  \gamnu][\gammu] \bigg)\,.
 \end{equation}
 After we substitute $l$ for $\alpha q$ and using
 $\Slaq/(\alpha \Slaq + \Slap - m_b) = 1/\alpha$
 from the on-shell condition, we arrive at the collinear divergence term $X_{(1a)}$ :
 \begin{equation}
 X_{(1a)} = -\frac{1}{2N_c}~\frac{2}{l^2 (l+u q)^2} \Big(\frac{\alpha+u}{\alpha}\Big)
 P(u)\,,
 \end{equation}
 where $P(u)$ is LO contribution of magnetic penguin operator
 defined by $P(u) = C_F [\Gamma^\mu_{q'}][\gamma_\mu]/(\ub m_b^2)$.
 Similarly, we can find the collinear divergences of diagrams (1d) and (1e):
  \begin{eqnarray}
 X_{(1d)} &=& -\frac{1}{2N_c}~\frac{2}{l^2 (l+u q)^2} \Big(\frac{\alpha+u}{\alpha}\Big)
 (-P(u+\alpha))\,, \\
  X_{(1e)} &=& \Big(\frac{1}{2N_c}+C_F\Big) \frac{2}{l^2 (l+u q)^2} \Big(\frac{\alpha+u}{\ub}\Big)(-P(u+\alpha))\,.
 \end{eqnarray}
 From the definition of $P(u)$, one can easily find the relation
 \begin{equation}
 P(u+\alpha) = \frac{\ub}{(\ub-\alpha)} P(u)\,,
 \end{equation}
 where we used on-shell condition in order to get $[\Gamma^\mu_{\qp}] = [\Gamma^\mu_{\pp}]$.
 Using this equation we can find that the collinear divergences proportional to the color factor $1/(2N_c)$ in
 $X_{(1a)}, X_{(1d)}$ and $X_{(1e)}$ are exactly canceled each other. The remaining collinear divergence
 term is proportional to $C_F$, and represented by
 \begin{equation}
 \label{eq-X1a1d1e}
 X_{(1a)+(1d)+(1e)} = C_F \frac{2}{l^2 (l+u q)^2} \Big(\frac{\alpha+u}{\alpha}\Big)
 (P(u)-P(u+\alpha)) \,.
 \end{equation}
 We also consider the collinear divergences of diagrams (1b), (1c) and (1f), and the calculation
 is similar to previous one. The result is
  \begin{equation}
  \label{eq-X1b1c1f}
 X_{(1b)+(1c)+(1f)} = C_F \frac{2}{l^2 (l+\ub q)^2} \Big(\frac{\alpha+\ub}{\alpha}\Big)
 (P(u)-P(u-\alpha)) \,.
 \end{equation}
 For the diagram (3), if we consider the projection operator eq.~(\ref{eq:proj}),
 we get the following collinear divergence term:
 \begin{equation}
 \label{eq-X3}
 X_{(3)} =  C_F \frac{2 \,\lt^2}{l^2 (l+u q)^2 (l - \ub q)^2} P(u+\alpha) \,.
 \end{equation}
 After we change $\alpha$ into $-\alpha$ in eq.~(\ref{eq-X1b1c1f}) and take apart the
 propagators in eq.~(\ref{eq-X3}),
 we obtain total collinear divergence term combining
 eqs.~(\ref{eq-X1a1d1e}), (\ref{eq-X1b1c1f}) and (\ref{eq-X3}):
 \begin{eqnarray}
 X_{\rm{tot}} &=& C_F \bigg(\frac{2(\alpha+u)}{\alpha} \frac{1}{l^2 (l+u q)^2}
 - \frac{2(\ub-\alpha)}{\alpha} \frac{1}{l^2 (l-\ub q)^2} \bigg) (P(u)-P(u+\alpha)) \nn \\
 &&~~~~~~~~ -C_F\,\frac{\lt^2}{q\cdot l} \bigg( \frac{1}{l^2 (l+u q)^2} - \frac{1}{l^2 (l-\ub q)^2} \bigg) P(u+\alpha)\,.
 \end{eqnarray}
 As this equation is same to eq.~(193) in ref.~\cite{Beneke:2000ry} except $T(u)$
 is changed into
 $P(u)$, this collinear divergence is to be canceled by one-loop correction of meson
 distribution amplitude. To be self-contained, here we show again how this cancelation
 take place. There is missing collinear divergence that comes from the self energy
 diagrams of upper side quark fields.  We find that this additional collinear divergence
 is given by
 \begin{equation}
 X_{\rm{add}} = - C_F \left(\frac{(\alpha+u)}{u} \frac{1}{l^2 (l+u q)^2}
 - \frac{(\alpha-\ub)}{\ub} \frac{1}{l^2 (l-\ub q)^2} \right) P(u)\,.
 \end{equation}
  We add $X_{\rm{add}}$ to $X_{\rm{tot}}$ and integrate them over $\beta$ and $\lt$ in order
  to express them as a convolution in $\alpha$. Using Cauchy's theorem for integration over
  $\beta$ and
  \begin{equation}
  \int \frac{d \lt^2}{\lt^2} =  2\ln \frac{\mu_{\rm{UV}}}{\mu_{\rm{IR}}} \,,
  \end{equation}
  we can find that $X_{\rm{tot}}+ X_{\rm{add}}$ is expressed by
 \begin{equation}
 X_{\rm{tot}}+ X_{\rm{add}} =
 C_F \frac{\alpha_s}{\pi} \ln \frac{\mu_{\rm{UV}}}{\mu_{\rm{IR}}} \int^1_0 dw P(w) V(w,u)\,.
 \end{equation}
 $V(w,u)$ is the Efremov-Radyushkin-Brodsky-Lepage (ERBL) kernel
 \cite{Efremov:1979qk,Lepage:1980fj}
 defined by
 \begin{equation}
 V(w,u) = \bigg[ \theta(u-w)\frac{w}{u}\bigg(1+\frac{1}{u-w}\bigg) + \theta(w-u)\frac{\bar{w}}{\bar{u}} \bigg( 1+\frac{1}{{\bar{u}-\bar{w}}}\bigg) \bigg]_+\,
 \end{equation}
 with the following definition
 \begin{equation}
 [f(w,u)]_+ \equiv f(w,u) - \delta(w-u) \int^1_0 dv f(v,u)\,.
 \end{equation}
 We note that the NLO contribution of light meson distribution amplitude is expressed by
 \begin{equation}
 \Phi^{(1)}_M(w) =  C_F \frac{\alpha_s}{\pi} \ln \frac{\mu_{\rm{UV}}}{\mu_{\rm{IR}}} \int^1_0 du V(w,u) \Phi^{(0)}_M(u)\,.
 \end{equation}
 Therefore, after the convolution with meson distribution amplitude, the total collinear
  divergence $X_{\rm{tot}}+X_{\rm{add}}$ in the matrix element is exactly canceled by NLO
  correction to the meson distribution amplitude in eq.~(\ref{eq:match}).

\begin{figure}[t]
\centerline{\epsfig{figure=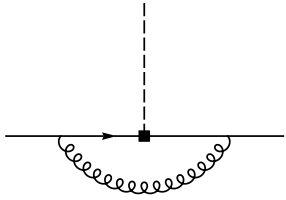, scale=0.3}}
\caption{NLO form-factor correction diagram.}
\label{fig:NLOFF}
\end{figure}

  Finally, we consider the collinear divergence that arises when the loop momentum $l$ is collinear to
 $\qp$. Diagrams (1c), (1d), (1f), (4) and (7d) have this collinear divergence. Without showing
 the details, it is found that the collinear divergences of diagrams (1c) and (1d) are canceled each
 other and so do the diagrams (1f) and (7d). The diagram (4) has both soft and collinear divergences.
 It is straightforward to show that the soft and collinear
  divergences of diagram (4) are exactly canceled by second term of eq.~(\ref{eq:match}) which is
  contributed by NLO form-factor correction diagram shown in figure~\ref{fig:NLOFF}.
  The UV divergence of NLO from-factor correction should be canceled by form-factor renormalization.

\subsection{ Renormalization - UV divergence cancelation }

\begin{figure}[t]
\centerline{\epsfig{figure=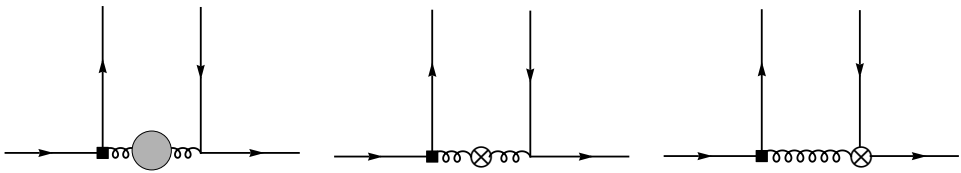, scale=0.3}}
\caption{Gluon self-energy diagrams and the counter terms.}
\label{fig:CT}
\end{figure}

 Now we add gluon self-energy diagrams as well as its counter term as shown in figure~\ref{fig:CT}.
The third diagram is the counter term for UV cancelation of diagrams (1c) and (1f).
 Other diagrams that cause UV divergence are diagrams (1a), (1e), (2), (5a), (5b), (6a-6d) and (7a-7c).
 These diagrams are same with those for calculating operator renormalization constant ${ Z}_{88}$
 except right side quark current. Therefore, all these UV divergences are naturally canceled by
 operator renormalization equation (\ref{eq:OPRen}).

 \section{Result and application}

  For the calculation, we first reduce the tensor integrals into scalar Feynman integrals.
 In order to reduce scalar Feynman integrals into the master integrals, we use the Laporta's algorithm \cite{Laporta:2001dd} with in-house Mathematica code which facilitates several reduction methods such as Passarino-Veltman reduction \cite{Passarino:1978jh},
  integration-by-part method \cite{Tkachov:1981wb,Chetyrkin:1981qh} and
  Lorentz-invariance method \cite{Gehrmann:1999as}.
  We mainly use the Mellin-Barnes representation to compute each master integral.
  Analytic results of the master integrals are shown in appendix.
   We explicitly separate the $1/\varepsilon$ and $1/\varepsilon^2$
 terms into IR divergence part and UV divergence part by investigating the origin of divergences  for each scalar Feynman integral.
  We confirm that all the IR divergences and UV divergences are canceled separately.
  Here we show the result of $T^{(0)}_{Q_8}$ and
 $T^{(1)}_{Q_8}$ (we attach subscription $Q_8$ in order to
 denote contribution of magnetic penguin operator) except the color factor $C_F/N_c$.
 The LO result
 $T^{(0)}_{Q_8} = - 2/\ub$ is same as in refs. \cite{Beneke:1999br,Beneke:2001ev}.

 The NLO result of $T^{(1)}_{Q_8}$ reads as
  \begin{eqnarray}
 u\bar{u}\,T^{(1)}_{Q_8} &=&
 \ln \Big(\frac{\mu }{m_b}\Big)
  \left(C_F \Big(8 u \ln \left(\bar{u}\right)-\frac{4u}{3}\Big)
  +\frac{8 u n_f}{3}-\frac{68 u}{3 N_c}\right) \nn \\
 &&  +i \pi  \Bigg(C_F \bigg(\Big(4u-\frac{4 \bar{u}}{u^2}\Big) \ln \left(\bar{u}\right)
  -\frac{8u}{3}-\frac{4}{u}+2\Bigg) \nn \\
 && ~~
  +\frac{1}{N_c} \bigg( \Big(2 u-\frac{2}{u^2}\Big) \ln\left(\bar{u}\right)
  -\frac{13 u}{3}-\frac{2}{u}-2 u \ln(u)-1\bigg) \Bigg) \nn \\
 &&+C_F \left(\Big(\,\frac{2 \bar{u}}{u^2}-2 u\Big) \ln^2\left(\bar{u}\right)
 +\Big(\,\frac{8u}{3}+\frac{4}{u}+2\Big) \ln \left(\bar{u}\right)-\frac{88 u}{9}+2\right) \nn\\
 && +\frac{1}{N_c} \left( \Big(\,\frac{2}{u}+1-3u\Big) J_1\left(\bar{u}\right)
     -\Big(2+\frac{2}{u}\Big)J_2(\bar{u})
     -6 u \text{Li}_2\left(\bar{u}\right) -2 u \text{Li}_2(u) \right. \nn \\
 && ~~
   +\Big(\frac{1}{u^2}+u\Big) \ln^2\left(\bar{u}\right) -u \ln ^2(u)
   -4 u \ln(\bar{u}) \ln(u)  \nn \\
 && ~~ \left.
   +\Big(\frac{19 u}{3}+\frac{2}{u}-1\Big) \ln \left(\bar{u}\right)
   +2 u \ln(u)
   +\frac{2 \pi ^2 u}{3}-\frac{242 u}{9}\right) \nn \\
 && + \,(n_f-2) \,g_8(0-i\epsilon,u) +  g_8(s_c-i\epsilon,u) + g_8(1,u)\,,
 \end{eqnarray}
 where $s_c = m_c^2/m_b^2~$. The definition of $g_8(s,u)$ is
  \begin{equation}
  g_8(s,u) = \left(\frac{8 u s}{3\bar{u}} +\frac{4}{3} \,u \right)
    J_1\left(\frac{\bar{u}}{s}\right) +\frac{16 u s}{3
   \bar{u}}-\frac{4}{3} u \ln (s)+\frac{20 u}{9}\,.
  \end{equation}
  The function $J_1(\ub)$ and $J_2(\ub)$ are defined by
   \begin{eqnarray}
  J_1(\ub) &=& \frac{1+y(\ub)}{1-y(\ub)} \ln (y(\ub))\,, \nn \\
  J_2(\ub) &=& \frac{\text{Li}_2(\ub)}{u} - \frac{\pi^2}{6u}
  -\int^1_0 d\xi \frac{\ln(1-\ub \,\xi (1-\xi))}{(1-\xi)(1-\ub \,\xi)}\,,
  \end{eqnarray}
  where
  \begin{equation}
  y(\ub) = \frac{\sqrt{4-\ub} - \sqrt{-\ub}}{\sqrt{4-\ub} + \sqrt{-\ub}} \,.
  \end{equation}
  In the case $s \to 0$, we find
  \begin{equation}
  g_8(0-i\epsilon,u)=  \frac{20 u}{9} -\frac{4}{3} u \ln (\ub) +i\pi \,\frac{4}{3}\, u\,.
  \end{equation}
  Even though $J_2(\ub)$ has integral form due to a non-trivial master integral,
 after convolution with meson distribution amplitude we could obtain complete
 analytic result of the hard-scattering amplitude. It should be emphasized that
 $u\bar{u}\, T^{(1)}_{Q_8}$ has no singularity except log-singularity
 as $u \to 0$ nor $u \to 1$, so that there is no end-point singularity for the
 convolution with meson distribution amplitude at leading twist.

  After convoluting the hard scattering kernels with meson distribution amplitude at
   leading twist, we are left with following formula:
  \begin{eqnarray}
  \label{eq:P8M1}
  P^{\rm{(0)}}_{8,M} &\equiv& \int^1_0 du\, T^{\rm{(0)}}_{Q_8} \Phi_M(u)
  = -6(1+\alpha^M_1 +\alpha^M_2)\,, \\
  \label{eq:P8M2}
   P^{\rm{(1)}}_{8,M} &\equiv& \int^1_0 du\, T^{\rm{(1)}}_{Q_8} \Phi_M(u)\nn\\
  &  = & 4 \pi  \text{Cl}_2\bigg(\frac{2
   \pi }{3}\bigg)-8 \text{Cl}_3\bigg(\frac{2 \pi}{3}\bigg)
   -\frac {332 \zeta (3)} {9} - \frac {10 \pi ^2} {9} + \frac {11
       \pi } {\sqrt {3}} - 123 \nn\\
  && +\left(16 \pi  \text{Cl}_2\bigg(\frac{\pi }{3}\bigg)-12 \pi
   \text{Cl}_2\bigg(\frac{2 \pi }{3}\bigg)-24 \text{Cl}_3\bigg(\frac{2 \pi
   }{3}\bigg) +\frac{892 \zeta (3)}{3}+\frac{38 \pi ^2}{3} \right. \nn \\
  &&\left. ~~~
   -49 \sqrt{3} \pi
   -\frac{4243}{9}\right) \alpha _1^{M}
   +\left(-96 \pi
   \text{Cl}_2\bigg(\frac{2 \pi }{3}\bigg)+192 \text{Cl}_3\bigg(\frac{2
   \pi }{3}\bigg) \right. \nn \\
  &&\left. ~~~
   -\frac{3464 \zeta (3)}{3}-60 \pi ^2+257 \sqrt{3} \pi
   +\frac{11011}{18}\right) \alpha _2^{M} + (n_f -2)G_{8}^M(0-i\epsilon) \nn \\
   &&  + G_{8}^M(s_c-i\epsilon) + G_{8}^M(1)
   +\left(-\frac{236 \alpha _1^{M}}{3}-\frac{308
   \alpha _2^{M}}{3}-36\right) \ln \Big(\frac{\mu }{m_b}\Big) \nn \\
   &&+i \pi
   \left(\left(52 \pi ^2-\frac{1747}{3}\right) \alpha
   _1^{M_2}+\left(\frac{6032}{3}-212 \pi ^2\right) \alpha _2^{M_2}-\frac{16
   \pi ^2}{3}+9\right)\,.
  \end{eqnarray}
  The function $G_8^M(s)$ is defined by
  \begin{eqnarray}
  G_8^M(s) &=& -24 s^2 \ln
   ^2\left(\tilde{y}(s)\right)+(4-40 s)\left(\frac{ 1+\tilde{y}(s)}{1-\tilde{y}(s)}\right) \ln \left(\tilde{y}(s)\right)-104 s-4 \ln (s)+\frac{38}{3} \nn \\
   &&+ \alpha_1^M \bigg(24 (8 s-9) s^2 \ln ^2\left(\tilde{y}(s)\right)-4(48
   s^2+34 s-1) \left(\frac{ 1+\tilde{y}(s)}{1-\tilde{y}(s)}\right)  \ln
   \left(\tilde{y}(s)\right)  \nn \\
  && ~~~
   -192 s^2-440 s-4 \ln (s)+18\bigg) \nn \\
   &&    +\alpha_2^M \bigg(-48 \left(45 s^2-40 s+18\right) s^2 \ln
   ^2\left(\tilde{y}(s)\right) \nn \\
  && ~~~
   + 4 \left(540 s^3-390 s^2-70 s+1\right) \left(\frac{ 1+\tilde{y}(s)}{1-\tilde{y}(s)}\right)   \ln \left(\tilde{y}(s)\right) \nn \\
  && ~~~
   +2160
   s^3-1740 s^2-1064 s-4 \ln (s)+21\Bigg)\,,
  \end{eqnarray}
 where
 \begin{equation}
 \tilde{y}(s) = \frac{\sqrt{1-4s} -1}{\sqrt{1-4s}+1} \,.
 \end{equation}
  In order to obtain the above equation, we use inverse binomial summation formula \cite{Davydychev:2003mv}.
 The Clausen function $\rm{Cl}_n(\theta)$ is defined in terms of poly-logarithm function \cite{Lewin:1981}
 \begin{equation}
 {\rm Cl}_n(x) = \left\{ \begin{array}{l}
 \frac{i}{2}\left( {\rm Li}_n \left(e^{-i\theta}\right) -
  {\rm Li}_n \left(e^{i\theta}\right) \right) ~~~ n{\rm ~is~even}
  \\
  \frac{1}{2}\left( {\rm Li}_n \left(e^{-i\theta}\right) +
  {\rm Li}_n \left(e^{i\theta}\right) \right)  ~~~ n{\rm ~is~odd} ~~.
   \end{array} \right.
 \end{equation}
 In eq.~(\ref{eq:P8M2}), we explicitly show the imaginary part in the last line.
 The other sources of imaginary part are $G^M_8(0-i\epsilon)$ and  $G^M_8(s_c-i\epsilon)$.
 These imaginary parts arise first at NLO (order  $\alpha_s^2$) for the magnetic penguin operator
 which causes strong CP phase in non-leptonic $B$ decays.

 Now we apply our result to $B \to K \pi$ decays. They are well studied in
 ref.~\cite{Beneke:2001ev} within the QCDF framework up to NLO. Here we follow the
 notation and the formula in ref.~\cite{Beneke:2001ev} for comparing our order $\alpha_s^2$
 result with their NLO one. The $B \to K \pi$ decay amplitudes
 ${\cal A} (B \to \pi K )$ without annihilation amplitudes are expressed by
 \begin{eqnarray}
 {\cal A} (B^- \to \pi^- \overline{K}^0 ) &=&
 \lambda_p \left[ \left( a^p_4 - \frac{1}{2} a^p_{10} \right)
 + r^K_\chi \left( a^p_6 - \frac{1}{2} a^p_8 \right) \right] A_{\pi K}, \nn \\
 -\sqrt{2}{\cal A} (B^- \to \pi^0 K^- ) &=&
 \left[ \lambda_u a_1 + \lambda_p ( a^p_4 + a^p_{10}) + \lambda_p \,r^K_\chi (a^p_6 +a^p_8)\right] A_{\pi K} \nn \\
  && +\left[ \lambda_u a_2 + \lambda_p \frac{3}{2} (-a_7+a_9) \right] A_{K\pi} ,\nn \\
  -{\cal A} (\overline{B}^0 \to \pi^+ K^- ) &=&
  \left[ \lambda_u a_1 + \lambda_p ( a^p_4 + a^p_{10}) + \lambda_p \,r^K_\chi (a^p_6 +a^p_8)\right] A_{\pi K}\,, \nn \\
  \sqrt{2} {\cal A} (\overline{B}^0 \to \pi^0 \overline{K}^0 ) &=&
    {\cal A} (B^- \to \pi^- \overline{K}^0 )+\sqrt{2}{\cal A} (B^- \to \pi^0 K^- ) \nn \\
    && -{\cal A} (\overline{B}^0 \to \pi^+ K^- )\,.
 \end{eqnarray}
 Here, $\lambda_p$ represents CKM factor defined by $\lambda_p = V_{pb} V^*_{ps}$, and
 the summation with $p = u, c$ is implicitly considered. It should be emphasized that
 the terms with $\lambda_u$ is highly CKM-suppressed as estimated by
  $|  V_{ub} V^*_{us} / V_{cb} V^*_{cs} | \approx 0.02 $.
 $A_{\pi K}$ and $A_{K \pi}$ contain all the hadronic parameters such as form-factor,
  meson decay constant and mass factor.
 The dimensionless coefficients $a_i$ represent the hard-scattering contribution
 of each operator combined with its Wilson coefficient. $r^K_\chi$ is chiral-enhancement
 factor defined by
 \begin{equation}
 r^K_\chi = \frac{ 2 m_K^2}{ \overline{m}_b \, (\overline{m}_q + \overline{m}_s)}\,,
 \end{equation}
 where $q = u$ for charged kaon and $q = d$ for neutral kaon.
  It is known that $a^c_4$ and
 $r^K_\chi a^c_6$ make dominant contribution in  $B \to K \pi$ decays among the $a_i$'s.
  Both come from the contribution of penguin operators. The magnetic penguin
  operator contributes to both $a^c_4$ and $r^K_\chi a^c_6$, but with different twist
  order of light-cone projection operator of pseudoscalar meson: leading twist for $a^c_4$
  and twist-3 for $r^K_\chi a^c_6$. Since, up to order $\alpha_s$, the twist-3  contribution of
  magnetic penguin operator is quite suppressed than the leading twist contribution,
  here we only consider NLO contribution of magnetic penguin operator to $a^c_4$ which
  requires leading twist projection.

  Without considering hard spectator scattering term (the term specified $a^c_{4,\rm{II}}$),
  we separate coefficient $a^c_{4,\rm{I}}$ (subscript I denotes the hard-scattering form-factor term)
  into three terms in order to compare the contributions of different sources:
  \begin{equation}
   a^c_{4,\rm{I}} = a^{c,V}_{4,\rm{I}} + a^{c,P}_{4,\rm{I}}+a^{c,P_8}_{4,\rm{I}}\,,
  \end{equation}
   where  $ a^{c,V}_{4,\rm{I}}$ denotes vertex correction term,
  $ a^{c,P}_{4,\rm{I}}$ represents penguin correction term and
  $a^{c,P_8}_{4,\rm{I}}$ is reserved for magnetic penguin operator contribution.
  We note that the three quantities are separately gauge-invariant, since they come from
  different gauge-invariant effective operators. Some care is needed for renormalization
  scheme dependence in each term.
  We see that the vertex correction starts from order unity while
  the penguin correction as well as the magnetic penguin correction start at $\alpha_s$,
  the vertex correction term $ a^{c,V}_{4,\rm{I}}$ makes dominant contribution.

  We show again the NLO expressions of each coefficient from ref.~\cite{Beneke:2001ev}
  including our result of NLO magnetic penguin operator contribution
  \begin{eqnarray}
  a^{c,V}_{4,\rm{I}} &=& C_4 + \frac{C_3}{N_c}\left(1+C_F \frac{\alpha_s}{4\pi}
  \bigg(12 \ln \frac{m_b}{\mu} - 18+ \int^1_0 du g(u) \Phi_K(u)\bigg) \right) \,,\nn \\
  a^{c,P}_{4,\rm{I}} &=& \frac{C_F}{N_c} \frac{\alpha_s}{4\pi} \Bigg(
  C_1 \bigg( \frac{4}{3} \ln \frac{m_b}{\mu} + \frac{2}{3} - G_K(s_c) \bigg)
  +C_3 \bigg(\frac{8}{3} \ln \frac{m_b}{\mu} + \frac{4}{3} - G_K(0) - G_K(1)\bigg) \nn\\
  &&+(C_4+C_6) \bigg( \frac{4 n_f}{3} \ln \frac{m_b}{\mu} - (n_f-2)G_K(0)-G_K(s_c)
  -G_K(1) \bigg) \Bigg) \,,\nn \\
  a^{c,P_8}_{4,\rm{I}} &=&  C^{\rm{eff}}_{8g} \frac{C_F}{N_c} \frac{\alpha_s}{4\pi}
  \left( P^{\rm{(0)}}_{8,K} + \left(\frac{\alpha_s}{4\pi}\right) P^{\rm{(1)}}_{8,K} \right) \,,
  \end{eqnarray}
  where the loop-effect functions $g(u)$, $G_K(s)$ are defined in ref.~\cite{Beneke:2001ev},
  and the  $P^{\rm{(0,1)}}_{8,K}$ can be read off from eqs.~(\ref{eq:P8M1}) and (\ref{eq:P8M2}).

 \begin{table}[t]
 \caption{ Input parameters. }
 \begin{tabular} {c|ccccc}
 \hline
 {\footnotesize Parameters              }~~&~~
 {\footnotesize $\Lambda_{\overline{\rm{MS}}}^{(5)}$ }~~&~~
 {\footnotesize $m_b(m_b)$              }~~&~~
 {\footnotesize $m_c(m_b)$              }~~&~~
 {\footnotesize $\alpha_1^K~(1\,\GeV)$  }~~&~~
 {\footnotesize $\alpha_2^K~(1\,\GeV)$  }\\
 {\footnotesize ~~ Values               }~~&~~
 {\footnotesize $225~ \MeV$             }~~&~~
 {\footnotesize $4.2~ \GeV$             }~~&~~
 {\footnotesize $1.3\pm 0.2 ~\GeV $     }~~&~~
 {\footnotesize $0.3 \pm 0.3$           }~~&~~
 {\footnotesize $0.1 \pm 0.3$           }\\
 \hline
 \end{tabular}
 \label{tab:InputParam}
 \end{table}

 For the numerical analysis, we compare the NLO contribution of magnetic penguin operator with
   its LO contribution and also with other penguin contribution and vertex contribution up to the
   NLO. We summarize input parameter values in table \ref{tab:InputParam}.
   We use two-loop running coupling constant and running quark mass with provided
   parameters. For the Gegenbauer moments, we use the values at the fixed scale with conservatively
   chosen errors that are consistent with several QCD sum rule results
   \cite{Khodjamirian:2004ga,Ball:2003sc}.
   With this preparation,
   we compute the values for $a^{c,P_8}_{4,\rm{I}}$ at the LO and NLO
   with three different scale values. The result is described in table \ref{tab:numerical}.
   For comparison, we also show numerical values of the $a^{c,V}_{4,\rm{I}}$ and $a^{c,P}_{4,\rm{I}}$ at NLO.
   The errors propagated from the uncertainties of
   Gegenbauer moments and charm quark mass are given in the round and square
   brackets respectively.
 \begin{table}[t]
 \caption{ Numerical result of coefficients $a^{c,P_8}_{4,\rm{I}}$ up to LO and NLO compared with
  NLO values of  $a^{c,V}_{4,\rm{I}}$ and $a^{c,P}_{4,\rm{I}}$ at three different
 scale values. The numbers in round brackets indicate maximal variation due to the
 uncertainty of Gegenbauer moments while the numbers in square brackets represent
 the variation from the charm mass uncertainty.  }
 \begin{tabular} {c|ccc|ccc}
 \hline
  &
  \multicolumn{3}{c|}{{\footnotesize Real part}} &
  \multicolumn{3}{c}{{\footnotesize Imaginary part}} \\
 {\footnotesize $\mu$             }&
 {\footnotesize $m_b/2$           }&
 {\footnotesize $m_b$             }&
 {\footnotesize $2m_b$            }&
 {\footnotesize $m_b/2$           }&
 {\footnotesize $m_b$             }&
 {\footnotesize $2m_b$            }\\
 \hline
 {\footnotesize $a^{c,V}_{4,\rm{I}}$  }&
 {\footnotesize $-0.046(1)$           }&
 {\footnotesize $-0.033$              }&
 {\footnotesize $-0.024$              }&
 {\footnotesize $-0.003(1)$           }&
 {\footnotesize $-0.001(1)$           }&
 {\footnotesize $-0.001$              }\\
 {\footnotesize  $a^{c,P}_{4,\rm{I}}$   }&
 {\footnotesize  $-0.003(2)[1]$         }&
 {\footnotesize  $-0.009(1)[1]$         }&
 {\footnotesize  $-0.014(1)[1]$         }&
 {\footnotesize  $-0.001(3)[4]$         }&
 {\footnotesize  $-0.003(2)[3]$         }&
 {\footnotesize  $-0.003(2)[3]$         }\\
 {\footnotesize  $a^{c,P_8}_{4,\rm{I}}$ (LO)  }&
 {\footnotesize  $0.014(2)$                   }&
 {\footnotesize  $0.009(2)$                   }&
 {\footnotesize  $0.007(1)$                   }&
 {\footnotesize  $0$                          }&
 {\footnotesize  $0$                          }&
 {\footnotesize  $0$                          }\\
 {\footnotesize  $a^{c,P_8}_{4,\rm{I}}$ (NLO) }&
 {\footnotesize  $0.019(4)$                   }&
 {\footnotesize  $0.014(3)$                   }&
 {\footnotesize  $0.010(2)$                   }&
 {\footnotesize  $0.007(2)$                   }&
 {\footnotesize  $0.004(1)$                   }&
 {\footnotesize  $0.002(1)$                   }\\
 \hline
 \end{tabular}
 \label{tab:numerical}
 \end{table}
 It should be noted that the contribution of magnetic penguin operator has opposite sign
 from the vertex and penguin correction, and compensates the other contribution.
 Specifically,  the real value of LO magnetic penguin contribution $a^{c,P_8}_{4,\rm{I}}$
 (order $\alpha_s$) is comparable to that of NLO penguin contribution
 $a^{c,P}_{4,\rm{I}}$ (order $\alpha_s$)  and both cancel each other at $\mu = m_b$.
 As can be seen, the NLO contribution of magnetic penguin operator is significant and can
 considerably reduce the absolute value of total coefficient $a^c_{4,I}$. Especially,
 for the imaginary part, $a^{c,P_8}_{4,\rm{I}}$(NLO) contribution is so strong that
 the contribution significantly reduces the total value.
 We emphasize that the large NLO contribution is mainly from the order $\alpha_s^2$
 QCD effect of the hard scattering term. We also comment that the contribution from
 the order $\alpha_s^2$ hard spectator scattering with QCD penguin contraction is very small,
 as can be seen in ref. \cite{Beneke:2006mk}.

 \section{Conclusions}

We computed the one-loop (order $\alpha_s^2$) contribution of magnetic penguin operator for
$B$ decays to light meson within the QCDF framework. It is explicitly shown that all the soft
and collinear divergences are canceled out. All the results are expressed in complete analytic
forms so that they can be easily applied to various non-leptonic $B$ decays. It turns out that
the NLO contribution of magnetic penguin operator is quite significant especially for
imaginary part in $B \to K\pi$ decays.
Therefore, the missing part of the order $\alpha_s^2$ correction to the
hard-scattering form-factor term for penguin amplitude is strongly required for precise estimates
of CP asymmetries.
It is interesting that the order $\alpha_s^2$ contribution of magnetic penguin operator
considerably reduces the absolute value of penguin amplitude coefficient $a_4^c$.
Combining future analysis for the order $\alpha_s^2$
penguin correction together with the current result of order $\alpha_s^2$ vertex
correction and hard-spectator scattering corrections, we could reach at the complete
order $\alpha_s^2$ accuracy for $B \to K\pi$ decays at the leading power of
$\Lambda_{\rm{QCD}}/m_b$.

\appendix
\section{Appendix: Analytic formulae of the master integrals}
\begin{figure}[t]
\centerline{\epsfig{figure=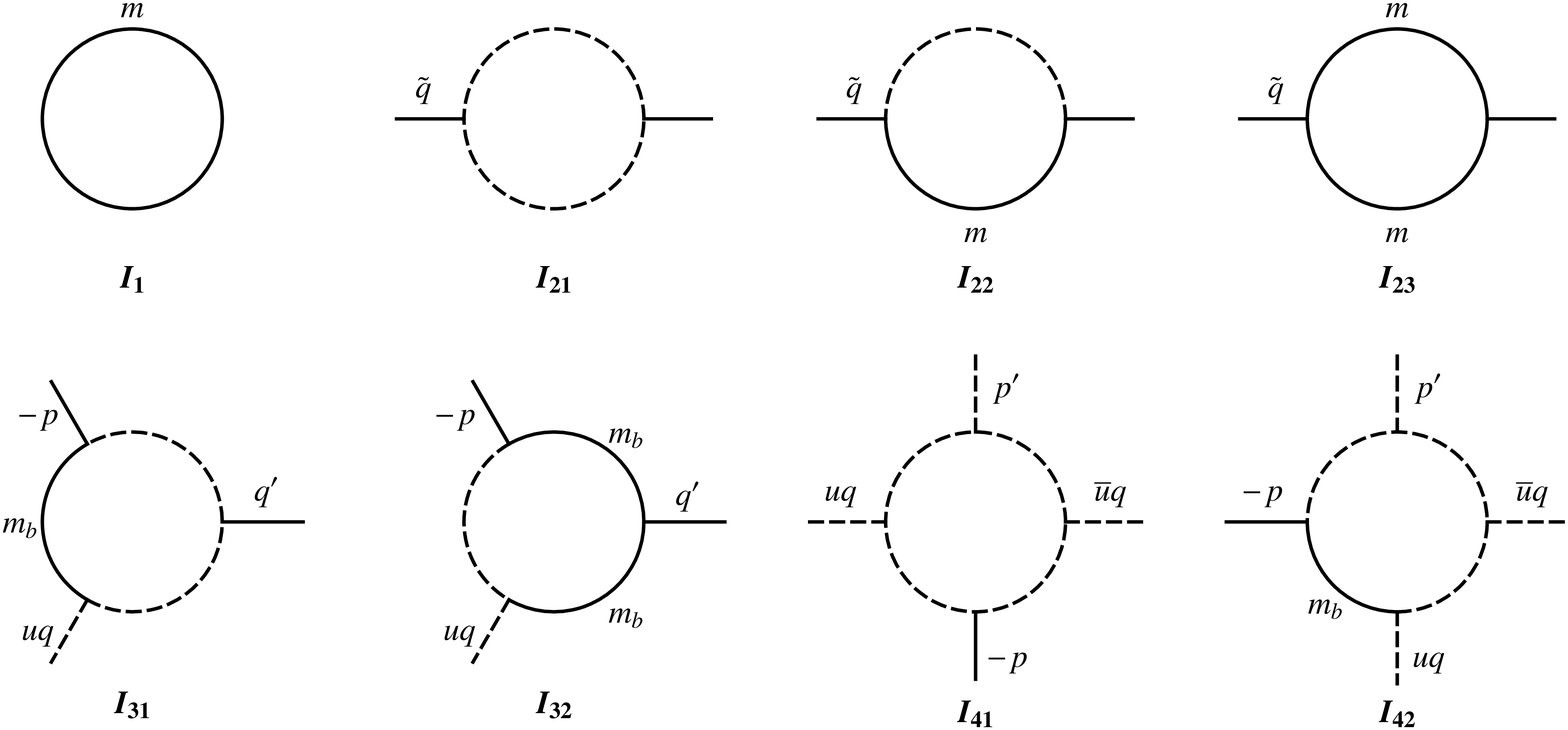, scale=0.28}}
\caption{Master Integrals.
Solid line represents `massive' for propagator whose mass is explicitly
shown in the diagram, and non-zero virtuality for external line. Dashed line implies massless
propagator or zero virtuality. All the momenta are understood to be ingoing.}
\label{fig:MIs}
\end{figure}

In this section, we summarize the result on the computation of the master integrals. We end up
with 8 master integrals which are displayed in figure~\ref{fig:MIs}. The calculation is based on
$\overline{\rm MS}$ renormalization scheme. Our convention for the master integral with propagators
${\cal P}_1,{\cal P}_2,...,{\cal P}_n$ is
\begin{equation}
(\mu e^{\gamma_E/2})^{4-d} \int \frac{d^d l}{i \pi^{d/2}}\, \frac{1}{{\cal P}_1{\cal P}_2\cdots{\cal P}_n}\,.
\end{equation}
\begin{eqnarray}
 I_1 &=& -m^2 \Big(\frac{\mu^2}{m^2}\Big)^{\varepsilon}( e^{\gamma_E \varepsilon})\Gamma(-1+\varepsilon\,) \,.\\
 I_{21} &=& \Big(\frac{\mu^2}{-{\tilde q}^2}+i 0\Big)^{\varepsilon}( e^{\gamma_E \varepsilon})
 \frac{\Gamma(1-\varepsilon)^2 \Gamma(\varepsilon)}{\Gamma(2-2\varepsilon)} \,. \\
 I_{22}({\tilde q}^2 < m^2) &=&  \Big(\frac{\mu^2}{m^2}\Big)^{\varepsilon} \bigg[
 2 -\frac{(x+1) \ln(x+1)}{x}+\frac{1}{\varepsilon } +\varepsilon\,\bigg(\left(\frac{1}{x}+1\right) \text{Li}_2(-x) \nn \\
 &&+\left(\frac{1}{x}+1\right) \ln
   ^2(x+1)-\frac{2 (x+1) \ln (x+1)}{x}+\frac{\pi ^2}{12}+4\bigg)
   + {\cal O}(\varepsilon^2)\, \bigg]\,, ~~\nn\\
   &&~~~ \textrm{ where}~~ x = -\frac{ {\tilde q}^2}{m^2}\,,\\
 I_{22}\,(\,{\tilde q}^2 =0) &=&  \Big(\frac{\mu^2}{m^2}\Big)^{\varepsilon} \bigg[
  \frac{1}{\varepsilon} + 1 + \varepsilon\,\left(1+\frac{\pi^2}{12}\right)+ {\cal O}(\varepsilon^2)
  \bigg]\,.
 \\
I_{23} &=& \Big(\frac{\mu^2}{m^2}\Big)^{\varepsilon} \bigg[
 \frac{1}{\varepsilon} + 2 + \frac{1+y(-x)}{1-y(-x)}\ln(y(-x)) + \cal{O}(\varepsilon)
\bigg]\,.  \\
I_{31} &=& \frac{1}{m_b^2} \Big(\frac{\mu^2}{m_b^2}\Big)^{\varepsilon} \bigg[
 -\frac{1}{2 u}\Big( (\ln(\bar u) - i\pi)^2 + \pi^2 \Big) + \cal{O}(\varepsilon)
\bigg]\,.  \\
I_{32} &=& \frac{1}{m_b^2} \Big(\frac{\mu^2}{m_b^2}\Big)^{\varepsilon} \bigg[
\frac{\text{Li}_2(\ub)}{u} - \frac{\pi^2}{6u}
  -\int^1_0 d\xi \frac{\ln(1-\ub \,\xi (1-\xi))}{(1-\xi)(1-\ub \,\xi)}  + \cal{O}(\varepsilon)
\bigg]\,.  \\
I_{41} &=& \frac{1}{m_b^4} \Big(-\frac{\mu^2}{m_b^2}+i 0 \Big)^{\varepsilon} \frac{1}{u\bar u} \bigg[
\frac {2} {\varepsilon ^2}-\frac {2 \ln \left (u \bar {u} \right) } {\varepsilon } \nn \\
&& -2 \text {Li}_2 (u) - 2 \text {Li}_2\left (\bar {u} \right)
+\ln^2\left (\frac {\bar {u}} {u} \right)+\frac {\pi ^2} {6} + \cal{O}(\varepsilon)
\bigg]\,. \\
I_{42} &=& \frac{1}{m_b^4} \Big(\frac{\mu^2}{m_b^2}\Big)^{\varepsilon}
 \frac{1}{\bar u} \bigg[
-\frac {3} {2 \,\varepsilon ^2} +\frac{ \ln(\bar{u})-i\pi}{\varepsilon} +\frac{13 \,\pi^2}{24}
+ \cal{O}(\varepsilon)
\bigg]\,.
\end{eqnarray}
We note that maintaining integral form in $I_{32}$ is convenient for convolution with meson
 distribution amplitude.

\section*{Acknowledgments}
Y.W.Y thanks KIAS Center for Advanced Computation for providing computing resources.
The work of C.S.K. was supported by the NRF grant
funded by the Korea government (MEST) (No. 2011-0027275) and (No. 2011-0017430).


\begin{thebibliography}{99}
\bibitem{:2008zza}
  S.~W.~Lin {\it et al.}  [The Belle Collaboration],
  {\it Difference in direct charge-parity violation between charged and neutral
  $B$ meson decays},
  Nature {\bf 452}, 332 (2008).

\bibitem{Aubert:2007hh}
  B.~Aubert {\it et al.}  [BABAR Collaboration],
  {\it Study of $B^0 \to \pi^0 \pi^0$, $B^\pm \to \pi^\pm \pi^0$, and $B^\pm \to
  K^\pm \pi^0$ decays, and isospin analysis of $B \to \pi \pi$ decays},
  Phys.\ Rev.\  D {\bf 76}, 091102 (2007)
  [arXiv:0707.2798].

\bibitem{Aubert:2007mj}
  B.~Aubert {\it et al.}  [BABAR Collaboration],
  {\it Observation of CP violation in $B^0 \to K^{+} \pi^{-}$ and $B^0 \to \pi^{+}
  \pi^{-}$},
  Phys.\ Rev.\ Lett.\  {\bf 99}, 021603 (2007)
  [arXiv:hep-ex/0703016].

\bibitem{Aaltonen:2011qt}
  T.~Aaltonen {\it et al.}  [CDF Collaboration],
  {\it Measurements of direct CP violating asymmetries in charmless decays of
  strange bottom mesons and bottom baryons},
  Phys.\ Rev.\ Lett.\  {\bf 106}, 181802 (2011)
  [arXiv:1103.5762].

\bibitem{Perazzini:2011kd}
  S.~Perazzini  [LHCb Collaboration],
  {\it Measurements of $A_{CP}(B^0\rightarrow K^+ \pi^-)$ and
  $A_{CP}(B_s\rightarrow \pi^+ K^-)$ at LHCb},
  [arXiv:1106.1197].

\bibitem{Beneke:1999br}
  M.~Beneke, G.~Buchalla, M.~Neubert and C.~T.~Sachrajda,
  {\it QCD factorization for $B \to \pi \pi$ decays: Strong phases and CP violation
  in the heavy quark limit},
  Phys.\ Rev.\ Lett.\  {\bf 83}, 1914 (1999)
  [arXiv:hep-ph/9905312].

\bibitem{Beneke:2000ry}
  M.~Beneke, G.~Buchalla, M.~Neubert and C.~T.~Sachrajda,
  {\it QCD factorization for exclusive, nonleptonic B meson decays: General
  arguments and the case of heavy light final states},
  Nucl.\ Phys.\  B {\bf 591}, 313 (2000)
  [arXiv:hep-ph/0006124].

\bibitem{Beneke:2001ev}
  M.~Beneke, G.~Buchalla, M.~Neubert and C.~T.~Sachrajda,
  {\it QCD factorization in $B \to \pi K, \pi \pi$ decays and extraction of
  Wolfenstein parameters},
  Nucl.\ Phys.\  B {\bf 606}, 245 (2001)
  [arXiv:hep-ph/0104110].

\bibitem{Beneke:2003zv}
  M.~Beneke and M.~Neubert,
  {\it QCD factorization for $B \to PP$ and $B \to PV$ decays},
  Nucl.\ Phys.\  B {\bf 675}, 333 (2003)
  [arXiv:hep-ph/0308039].

\bibitem{Bell:2007tv}
  G.~Bell,
  {\it NNLO vertex corrections in charmless hadronic $B$ decays: Imaginary part},
  Nucl.\ Phys.\  B {\bf 795}, 1 (2008)
  [arXiv:0705.3127].

\bibitem{Bell:2009nk}
  G.~Bell,
  {NNLO vertex corrections in charmless hadronic $B$ decays: Real part},
  Nucl.\ Phys.\  B {\bf 822}, 172 (2009)
  [arXiv:0902.1915].

\bibitem{Beneke:2009ek}
  M.~Beneke, T.~Huber and X.~Q.~Li,
  {\it NNLO vertex corrections to non-leptonic $B$ decays: Tree amplitudes},
  Nucl.\ Phys.\  B {\bf 832}, 109 (2010)
  [arXiv:0911.3655].



\bibitem{Beneke:2005vv}
  M.~Beneke and S.~Jager,
  {\it Spectator scattering at NLO in non-leptonic $b$ decays: Tree amplitudes},
  Nucl.\ Phys.\  B {\bf 751}, 160 (2006)
  [arXiv:hep-ph/0512351].

\bibitem{Pilipp:2007mg}
  V.~Pilipp,
  {\it Hard spectator interactions in $B \to \pi \pi$ at order $\alpha_s^2$},
  Nucl.\ Phys.\  B {\bf 794}, 154 (2008)
  [arXiv:0709.3214].

\bibitem{Kivel:2006xc}
  N.~Kivel,
  {\it Radiative corrections to hard spectator scattering in $B \to \pi \pi$
  decays},
  JHEP {\bf 0705}, 019 (2007)
  [arXiv:hep-ph/0608291].

\bibitem{Beneke:2006mk}
  M.~Beneke and S.~Jager,
  {\it Spectator scattering at NLO in non-leptonic $B$ decays: Leading penguin
  amplitudes},
  Nucl.\ Phys.\  B {\bf 768}, 51 (2007)
  [arXiv:hep-ph/0610322].

\bibitem{Jain:2007dy}
  A.~Jain, I.~Z.~Rothstein and I.~W.~Stewart,
  {\it Penguin loops for nonleptonic $B$-decays in the standard model: Is there a
  penguin puzzle?},
  [arXiv:0706.3399].

\bibitem{Buchalla:1995vs}
  G.~Buchalla, A.~J.~Buras and M.~E.~Lautenbacher,
  {\it Weak decays beyond leading logarithms},
  Rev.\ Mod.\ Phys.\  {\bf 68}, 1125 (1996)
  [arXiv:hep-ph/9512380].

\bibitem{Grinstein:1990tj}
  B.~Grinstein, R.~P.~Springer and M.~B.~Wise,
  {\it Strong interaction effects in weak radiative anti-$B$ meson decay},
  Nucl.\ Phys.\  B {\bf 339}, 269 (1990).


\bibitem{Chetyrkin:1996vx}
  K.~G.~Chetyrkin, M.~Misiak and M.~Munz,
  {\it Weak radiative $B$ meson decay beyond leading logarithms},
  Phys.\ Lett.\  B {\bf 400}, 206 (1997)
  [Erratum-ibid.\  B {\bf 425}, 414 (1998)]
  [arXiv:hep-ph/9612313].

\bibitem{Greub:2000sy}
  C.~Greub and P.~Liniger,
  {\it Calculation of next-to-leading QCD corrections to $b \to sg$},
  Phys.\ Rev.\  D {\bf 63}, 054025 (2001)
  [arXiv:hep-ph/0009144].

\bibitem{Floratos:1977au}
  E.~G.~Floratos, D.~A.~Ross and C.~T.~Sachrajda,
  {\it Higher order effects in asymptotically free gauge theories: The anomalous
  dimensions of wilson operators},
  Nucl.\ Phys.\  B {\bf 129}, 66 (1977)
  [Erratum-ibid.\  B {\bf 139}, 545 (1978)].

\bibitem{GonzalezArroyo:1979df}
  A.~Gonzalez-Arroyo, C.~Lopez and F.~J.~Yndurain,
  {\it Second order contributions to the structure functions in deep inelastic
  scattering. 1. Theoretical calculations},
  Nucl.\ Phys.\  B {\bf 153}, 161 (1979).

\bibitem{Mueller:1993hg}
  D.~Mueller,
  {\it Conformal constraints and the evolution of the nonsinglet meson
  distribution amplitude},
  Phys.\ Rev.\  D {\bf 49}, 2525 (1994).

\bibitem{Efremov:1979qk}
  A.~V.~Efremov and A.~V.~Radyushkin,
  {\it Factorization and asymptotical behavior of pion form-factor in QCD},
  Phys.\ Lett.\  B {\bf 94}, 245 (1980).

\bibitem{Lepage:1980fj}
  G.~P.~Lepage and S.~J.~Brodsky,
  {\it Exclusive processes in perturbative quantum chromodynamics},
  Phys.\ Rev.\  D {\bf 22}, 2157 (1980).

\bibitem{Laporta:2001dd}
  S.~Laporta,
  {\it High precision calculation of multiloop Feynman integrals by difference
  equations},
  Int.\ J.\ Mod.\ Phys.\  A {\bf 15}, 5087 (2000)
  [arXiv:hep-ph/0102033].


\bibitem{Passarino:1978jh}
  G.~Passarino and M.~J.~G.~Veltman,
  {\it One loop corrections for $e^+ e^-$ annihilation into $\mu^+ \mu^-$
  in the Weinberg model},
  Nucl.\ Phys.\  B {\bf 160}, 151 (1979).

\bibitem{Tkachov:1981wb}
  F.~V.~Tkachov,
  {\it A theorem on analytical calculability of four loop renormalization group
  functions},
  Phys.\ Lett.\  B {\bf 100} (1981) 65.

\bibitem{Chetyrkin:1981qh}
  K.~G.~Chetyrkin and F.~V.~Tkachov,
  {\it Integration by parts: The algorithm to calculate beta functions in 4 loops},
  Nucl.\ Phys.\  B {\bf 192} (1981) 159.

\bibitem{Gehrmann:1999as}
  T.~Gehrmann and E.~Remiddi,
  {\it Differential equations for two loop four point functions},
  Nucl.\ Phys.\  B {\bf 580}, 485 (2000)
  [arXiv:hep-ph/9912329].



\bibitem{Davydychev:2003mv}
  A.~I.~Davydychev and M.~Y.~Kalmykov,
  {\it Massive Feynman diagrams and inverse binomial sums},
  Nucl.\ Phys.\  B {\bf 699}, 3 (2004)
  [arXiv:hep-th/0303162].

\bibitem{Lewin:1981}
 L. Lewin, {\it Polylogarithms and associated functions}, North-Holland, Amsterdam, (1981).


\bibitem{Khodjamirian:2004ga}
  A.~Khodjamirian, T.~Mannel and M.~Melcher,
  {\it Kaon distribution amplitude from QCD sum rules},
  Phys.\ Rev.\  D {\bf 70}, 094002 (2004)
  [arXiv:hep-ph/0407226].

\bibitem{Ball:2003sc}
  P.~Ball and M.~Boglione,
  {\it SU(3) breaking in $K$ and $K^*$ distribution amplitudes},
  Phys.\ Rev.\  D {\bf 68}, 094006 (2003)
  [arXiv:hep-ph/0307337].


\end{thebibliography}
\end{document}